\documentclass{ws-procs9x6}

\begin{document}

\title{$B \to K^* \gamma$ vs $B \to \rho \gamma$ and  $|V_{\rm td}/V_{\rm ts}|$}

\author{Patricia Ball$^*$ and Roman Zwicky$^\dagger$}

\address{IPPP, Department of Physics, University of Durham, \\
Durham DH1 3LE, UK\\
$^*$E-mail: patricia.ball@durham.ac.uk \quad 
$^\dagger$roman.zwicky@durham.ac.uk}

\begin{abstract}
We determine $|V_{\rm td}/V_{\rm ts}| = 0.192  \pm 0.016 ({\rm exp}) \pm 0.014({\rm th} )$   from $B \to (K^*,\rho) \gamma$ as measured by the B-factories Babar and Belle. 
\\[0.5cm]
{\bf This version differs from the original proceedings  by the inclusion of 
the new BaBar measurement of \boldmath $ 
B \to \rho(\omega) \gamma$ presented at ICHEP 2006, which significantly
shifts the results for $|V_{td}/V_{ts}|$.}
\end{abstract}

\bodymatter

\section{Introduction}\label{sec1}
Decays like $B \to (K^*,\rho)\gamma$ are the exclusive counterparts of the famous flavour changing neutral
current (FCNC) decays $b \to (s,d) \gamma$. It is well known that FCNC
are absent in the Standard Model (SM) at the tree-level and further
suppressed at the loop-level by the GIM-mechanism. Therefore they
represent a sensitive probe for new physics (NP) and are widely
studied in the literature. Moreover the weak transition
involves the CKM-matrix elements $V_{t(d,s)}$. 
 
The decays are described in the language of effective Hamiltonians
$H_{\rm eff} = \sum_i C_i(\mu)O_i(\mu)$ where the Wilson coefficients $C_i$ are matched to the SM or some beyond SM framework at the electroweak scale. As a next step 
the renormalization group is employed to evolve the coefficients 
down to a scale $\mu$ where the actual decay of the B-meson 
takes place. This procedure has been performed in QCD to NLO \cite{short} 
and the NNLO calculation are expected to be finished soon.
As a final step the matrix elements of the operators $O_i$ have
to be taken, which is a notoriously difficult task due to bound-state effects.  The matrix elements are parametrized in terms of form factors
and it is the aim of this text to report on the reduction of the uncertainty
of the relevant form factors in $B \to K^* \gamma$ and $B \to \rho \gamma$ \cite{BZ06b}.

On the experimental side the mode $B \to K^* \gamma$ has been measured as early as
1993 by CLEO. Its inclusive counterpart $b\ \to s \gamma$ is in very good agreement with NP calculations and gives stringent constraints on physics beyond the SM.
The decay $B \to \rho \gamma$ has recently been measured  by Belle \cite{Belle} and
very recently by Babar \cite{Babar}.

\section{Reduced uncertainty in the ratio of branching ratios}
The effective Hamiltonian for the $b \to D \gamma$ transition reads $(D=d,s)$
\begin{equation*}
\label{eq:heff}
 H^{b \to D}_{\rm eff} =  \frac{4 G_F}{\sqrt{2}}\Big( \sum_{U=u,c} \lambda_U^{(D)} 
\Big[ C_1 O_1^{U,(D)} + C_2 O_2^{U,D} \Big]      +      (-\lambda_t^{(D)}) \sum_{i=3..8} C_i O_i    \Big) 
\end{equation*}
where $O_{1,2}$ are current-current operators,
 $O_{3 ..6}$ penguin operators and $O_{7,8}$ are the electric and magnetic dipole operators. The most important ones are 
\begin{alignat}{2}
\label{eq:O7}
& O_7       &=&   \frac{e m_b}{32 \pi^2} ({\bar D}  \sigma_{\mu\nu}F^{\mu\nu} (1+\gamma_5)  b) \quad ,\\[0.2cm]
\label{eq:O2}
& O_2^{U,D} &=& \frac{1}{4} ({\bar D }  \gamma_\mu(1\!-\!\gamma_5) U) (\bar U  \gamma^\mu(1\!-\!\gamma_5) b) \quad ,
\end{alignat}
where $U=u,c$ as above and the  $\lambda$'s are CKM factors, 
e.g. $\lambda_t^s = V_{\rm tb}V_{\rm ts}^*$. \\
It has recently been shown in QCD factorization that at leading order
in $1/m_b$ the following factorization formula applies \cite{BVg04}
\begin{eqnarray}\label{1}
\langle V\gamma|O_i| B\rangle =
\left[ T_1^{B\to V}(0)\, T^I_{i} +
\int^1_0 d\xi\, du\, T^{II}_i(\xi,u)\, \phi_B(\xi)\, \phi_{V;\perp}(v)\right]
\cdot\epsilon\, ,
\end{eqnarray}
where $\epsilon$ is the photon polarisation 4-vector, $O_i$ is one of
the operators in the  effective Hamiltonian,
$T_1^{B\to V}$ is a $B\to V$ transition form factor,
and $\phi_B$, $\phi_{V;\perp}$ 
are leading-twist light-cone distribution amplitudes (DA)
of the $B$ meson and the vector meson $V$, respectively.
These quantities are universal non-perturbative objects and
describe the long-distance dynamics of the matrix elements, which
is factorised from the perturbative short-distance interactions
included in the hard-scattering kernels $T^I_{i}$ and $T^{II}_i$.
The amplitude reads
\begin{equation*}
A(B \to (K^*,\rho) \gamma)=  \Big[ \frac{4 G_F}{\sqrt{2}}V_{\rm tb}V_{\rm t(s,d)}^* \Big] \, a_7(K^*,\rho) 
\underbrace{ \langle V \gamma |  O_7 | B  \rangle}_{\sim T_1^{B \to V}(0)} + O(1/m_b)
\end{equation*}
with $a_7 = C_7 + O(\alpha_s)$ . The most important $1/m_b$ corrections are due to the
operator $O_{2}$ \eqref{eq:O2}, which come with a numerically enhanced Wilson coefficient $C_2(m_b) \sim 3|C_7(m_b)|$. A further hierarchy is set by CKM factors, c.f. Tab. \ref{tbl1}, which implies that
\begin{table}
\tbl{CKM-hierarchy for operators $O_{2(1)}$. 
$\lambda \sim 0.22$ is a Wolfenstein-parameter.}
{\begin{tabular}{@{}ccc@{}}\toprule
$\lambda^{D}_U$ & $U = u$ &  $U=c$  \\\colrule
$\rho\hphantom{^*}: D=d$  \quad & $ \lambda^3$ & $\lambda^3$ \\\colrule
$K^*:D=s$    &  $\lambda^4$ & $\lambda^2$  \\\botrule
\end{tabular}}
%\begin{tabnote}
%$\^{\text a}$ Sample table footnote.\\
%\end{tabnote}
\label{tbl1}
\end{table}
for the $B \to K^*\gamma$ transition solely the current-current operator with a $c$-quark pair is numerically relevant. This contribution
has been estimated a long time ago \cite{KRSW97} and is considered to be under reasonable control. Moreover, assuming three generations,  $|V_{ts}|=|V_{cb}|(1+O(\lambda^2))$, and $|V_{cb}|$ is known with a precision of $2\%$  \cite{Vcb}. Therefore, the major unknown quantity is the penguin form factor $T_1^{B \to K^*}(0)$. The value quoted by the authors in $^\dagger$,\cite{BVg04} compares well to an update of the form factors from Light Cone Sum Rules (LCSR) \cite{BZ04b} 
\begin{equation*}
{T_1}^{B \to K^*}  (0)^{\dagger} = 0.28 \pm 0.02 \leftrightarrow
{T_1}^{B \to K^*} _{\, \rm LCSR}(0) = 0.31 \pm 0.04
\end{equation*}
due to recent progress in the Kaon distribution amplitude to be discussed below.
For the $B \to \rho\gamma$ transition things are more complicated because there also the current-current operators with an $u$-quark 
contribute which include annihilation contributions and an $u$-quark 
loop which has so far only been addressed by model dependent calculations, which makes the extraction of $|V_{\rm td}|$ by itself difficult.

On the other hand, in the ratio of branching ratios
\begin{equation}\label{Brat}
\frac{{\cal B}(B\to \rho \gamma)}{{\cal B}(B\to K^*\gamma)} =
\left|\frac{V_{td}}{V_{ts}}\right|^2
\left(\frac{m_B^2-m_{\rho}^2}{m_B^2-m_{K^*}^2}\right)^3
\underbrace{\left( \frac{T_1^{\rho}(0)}{T_1^{K^*}(0)}\right)^2}_{\equiv \xi^{-2}} \left|
  \frac{a_7^c(\rho)}{a_7^c(K^*)}\right|^2 \left[ 1\! +\!
  \Delta R\right] \quad ,
\end{equation}
the $1/m_b$-corrections $\Delta R$  are somewhat accidently  CKM-suppressed.
The parameters
$|a_7^c(\rho\gamma)|$ and $|a_7^c(K^*\gamma)|$ are almost exactly
equal, so we set $|a_7^c(\rho\gamma)/$ $a_7^c(K^*\gamma)|=1$. Non-factorizable contributions are unlikely 
to change this ratio significantly \cite{KRSW97},\cite{BZ06c}.
In order to compensate for relatively poor statistics 
an averaging is performed over isospin and we completely neglect the 
difference between the $\rho$ and the $\omega$ meson. Then
\begin{equation*}
\label{eq:deltaR}
\Delta R_{\pm,0}  \simeq  {\rm Re}\,(\delta a_\pm +  \delta a_0)  f_{\rm CKM}
 + \frac{1}{2}\left( |\delta a_\pm|^2 + |\delta a_0|^2\right)g_{\rm CKM}  \quad ,
\end{equation*}
where  $\delta a_{0,\pm}= a_7^u(\rho^{0,\pm}\gamma)/a_7^c(\rho^{0,\pm}\gamma)-1$. The almost-equal sign indicates that other types of corrections are Cabibbo-suppressed and the
subscript is a reminder of the isospin averaging. 
The CKM factors are given by
\begin{eqnarray*}
f_{\rm CKM} &=& \frac{R_b^2 - R_b \cos\gamma}{1-2 R_b \cos\gamma + R_b^2} = 0.07 \pm 0.12  \\
g_{\rm CKM} &=&   \frac{R_b^2}{1-2 R_b \cos\gamma + R_b^2}  
= 0.23 \pm 0.07 \quad ,
\end{eqnarray*}
with $\gamma$ being one angle of  the unitarity triangle and $R_b \equiv 
(1- \lambda^2/2 )\frac{1}{\lambda} \left|V_{ub}/V_{cb} \right|$ one of its sides. The uncertainties are obtained by using 
$\gamma  = (71 \pm 16)^\circ$, the value of $R_b$ has
only a minor impact as compared to the former. 

\subsection{\boldmath The form factor ratio $T_1^{B\to K^*}/T_1^{B\to
    \rho}$}
The penguin form factor is defined by the following matrix element
\begin{equation}
\langle V(p)|\bar D \sigma_{\mu\nu} q_{\nu}(1 \! + \! \gamma_5) b|B(p_B) \rangle  = i \epsilon_{\mu\nu\rho\sigma} \varepsilon^{*\nu} p_B^\rho p^\sigma
2 T_1^{B \to V}(0) 
\end{equation}
at zero momentum transfer $q^2 = (p_B-p)^2 =0$. In reference \cite{BZ04b}
this form factor was calculated using LCSR including radiative corrections up to next-to-leading twist. The uncertainty of the ratio 
\begin{equation}
\label{eq:T1ratio}
\xi \equiv \frac{T_1^{B \to K^*}(0)}{T_1^{B \to \rho}(0)} \quad ,
\end{equation}
we are aiming at, is considerably smaller than that of the individual form factors themselves
as many uncertainties tend to cancel each other,
e.g. the dependence on the $b$-quark mass and the normalization
through the B-meson decay constant $f_B$.  The form factor ratio 
$\xi$ is basically a measure of $SU(3)$-breaking.

In reference \cite{BZ06b} we have 
further improved this calculation w.r.t. \cite{BZ04b} by including updated $SU(3)$-breaking in the twist-2 parameters, complete account of $SU(3)$-breaking in the twist 3 and 4 DA  and
we have tested the sensitivity of the ratio to different models of DA.
It was found that solely the SU(3) breaking of the leading twist-2 DA has
a major impact on the numerical value of $\xi$ and we will therefore shortly discuss its status.

\subsection{SU(3)-breaking of distribution amplitudes}
The most important contribution to $T_1^{B \to V}$
comes from the leading-twist transverse DA
\begin{equation*}
 \langle 0 |\bar q(z) \sigma_{\mu\nu} D (-z)
  |V (q,\lambda) \rangle  = 
  i(e^{(\lambda)}_\mu q_\nu -e^{(\lambda)}_\nu q_\mu)
  f_V^\perp(\mu) \int_0^1 du\, e^{i\xi q \cdot z} \phi_V^\perp(u),
\end{equation*}
and we refer  to \cite{BZ05b},\cite{BZ06a} for further details and references.
The normalization of the transverse DA cannot be directly accessed
by experiment unlike the longitudinal DA. We therefore rely on
theoretical calculations. The updated QCD sum rule calculations for these
quantities are \cite{BZ06b}
\begin{equation}
\label{eq:fT}
f_\rho^\perp(1\,{\rm GeV}) = (0.165\pm 0.009)\,{\rm GeV}  ,
f_{K^*}^\perp(1\,{\rm GeV}) = (0.185\pm 0.010)\,{\rm GeV}  .
\end{equation}
The uncertainties of these quantities are rather large by themselves as
compared to the longitudinal decay constants
$$f_\rho^\parallel = (0.209\pm 0.002)\,{\rm GeV},\qquad
 f_{K^*}^\parallel = (0.217\pm
0.005)\, {\rm GeV}.$$
The uncertainties in \eqref{eq:fT} dominate the uncertainty  in the final extraction of $|V_{\rm td}/V_{\rm ts}|$. Lattice QCD \cite{becirevic}
has up to now only produced ratios of the longitudinal to transverse decay constants. 
The ratios of decay constants in 
Sum Rules \cite{BZ06b}  and Lattice-QCD calculations agree  reasonably well.
Nevertheless further efforts in this direction would be highly desirable.

The other important quantity for the tensor ratio \eqref{eq:T1ratio} is the first
Gegenbauer $a_1^\perp(V)$  moment in the conformal expansion of the DA, 
\begin{equation*}
\phi_\perp(u) = 6u\bar u\left(1 + \sum_{n \geq 1} a_n^\perp(V) C_n^{3/2}(2u-1)\right) \quad .
\end{equation*}
The first Gegenbauer  moment $a_1$ is zero for particles with a definite G-parity as the 
$\rho$-meson.
Moreover it is a measure for the average momentum of the strange quark as compared to the
light quark in the Kaon and it is therefore expected to be positive on 
intuitive grounds. 
A positive value was found in \cite{Russians} a long time ago, but when radiative corrections
were evaluated and a sign mistake corrected, the value of $a_1$ became  negative 
\cite{elena}. In the following year two papers appeared \cite{alex},\cite{lenz} which criticized the instability of the non-diagonal sum rule used in \cite{Russians},\cite{elena}.
One paper used an exact operator relation \cite{lenz}  and the other one used stable diagonal sum rules \cite{alex}. Both analyses were completed in reference \cite{BZ05b} and \cite{BZ06a}.
It is found that the
diagonal sum rules are numerically superior to the operator method and the final values for $a_1$ are (at $\mu = 1\,{\rm GeV}$)  \cite{BZ05b},\cite{BZ06a},\cite{BBL}
$$
a_1(K) = 0.06\pm 0.03, \quad a_1^\parallel(K^*) = 0.03\pm 0.02,\quad  
a_1^\perp(K^*) = 0.04\pm 0.03.
$$

With these values we obtained the following result \cite{BZ06b}
\begin{eqnarray}
\xi = \frac{T_1^{B\to K^*}(0)}{T_1^{B\to \rho}(0)} &=& 1.17 \pm
0.08(f^\perp_{\rho,K^*}) \pm 0.03(a_1)
\pm 0.02(a_2) \pm 0.02(\mbox{t-3,4})\nonumber\\
&&{} \pm 0.01 (\mbox{sum-rule
  param., $m_b$, t-2 and -4 models})\nonumber\\
&= &1.17\pm 0.09\,,\label{resxi}
\end{eqnarray}
where the total uncertainty of $\pm 0.09$ is
obtained by adding the individual terms in quadrature.

\subsection{Extraction  of $|V_{\rm td}/V_{\rm ts}|$}
The Belle collaboration
has measured the quantity 
$$
R_{\rm exp}  \equiv 
\frac{\overline{\cal B}(B\to (\rho,\omega)\gamma)}{\overline{\cal B}(B\to
  K^*\gamma)}\,,
$$
where $\overline{\cal B}(B\to (\rho,\omega)\gamma)$ is defined as the
CP-average of 
$${\cal B}(B\to (\rho,\omega)\gamma) = \frac{1}{2}\left\{ {\cal B}(B^+\to
\rho^+\gamma) + \frac{\tau_{B^+}}{\tau_{B^0}}\left[ {\cal B}(B^0\to
\rho^0\gamma) + {\cal B}(B^0\to\omega\gamma)\right]\right\},
$$
and $\overline{\cal B}(B\to K^*\gamma)$ is the isospin- and
  CP-averaged branching ratio of the $B\to K^*\gamma$ channels. 
HFAG averages Belle \cite{Belle} and Babar \cite{Babar}  results into  \cite{HFAG}
\begin{equation}\label{RexHFAG}
R_{\rm exp}^{\rm HFAG}= 0.028\pm 0.005\,.
\end{equation}
This has to be compared to the theoretical prediction
\begin{equation}
\label{eq:Rth}
R_{\rm th} = 
\left|\frac{V_{td}}{V_{ts}}\right|^2 \left[0.75\pm 0.11(\xi)\pm
0.02(a_7^{u,c},\gamma,R_b)\right], \\
\end{equation}
where in \eqref{eq:deltaR} the annihilation contributions  have been evaluated 
in LCSR as well as QCD factorization and although the individual results differ slightly it has an absolutely negligible effect on the value and the uncertainty of the ratio due to the
accidental CKM-supression \cite{BZ06b}.
Equating we extract
\begin{equation}
\left|\frac{V_{td}}{V_{ts}}\right|_{B\to V\gamma}^{\rm HFAG}  =   
0.192  \pm 0.016 ({\rm exp}) \pm 0.014({\rm th})\,,\label{Rtrad}
\end{equation}
which is our final result. It is interesting to compare this result to 
two other determinations of $|V_{\rm td}/V_{\rm ts}|$.
First in the SM the ratio is related to other
CKM-parameters by unitarity and allows an estimate from the CKM-fit
\begin{eqnarray*}
\label{eq:fit}
\left|\frac{V_{td}}{V_{ts}}\right|_{\rm SM}  = 
\lambda (1+ R_b^2 - 2 R_b\cos\gamma)^{1/2}
&=& 0.216\pm 0.029\, \label{RtSM}  .
\end{eqnarray*}
It is also interesting to note that for a typical LHCb uncertainty of the angle $\gamma$ of about $4\%$ the above uncertainty would drop to about $\sim 0.007$.
Moreover this determination can be compared to that from
$B_{(d,s)}$-oscillations which have been measured by CDF;
they quote the following value \cite{CDF}
\begin{equation}
\label{eq:BsBd}
\left|\frac{V_{td}}{V_{ts}}\right|_{\rm CDF,\Delta M_s} = 0.206 \pm 0.0007 ({\rm exp}) ^{+0.008}_{-0.006} ({\rm theo}) \quad  .
\end{equation}

\section{Conclusions}
We have calculated the ratio of tensor form factors determining the decays $B \to (K^*,\rho) \gamma$ as $\xi = 1.17 \pm 0.09$. Its uncertainty is dominated by the uncertainty of transversal decay constants. We then have calculated the isospin averaged ratio of branching ratios \eqref{eq:Rth}, including a preliminary estimate of power correction to be completed
in \cite{BZ06c}. From this we have obtained  $|V_{\rm td}/V_{\rm ts}|=0.179  \pm 0.022 ({\rm exp}) \pm 0.014({\rm th})$. This value is consistent, within uncertainties, with the value from the CKM-fit \eqref{eq:fit} and the value from $B_{(d,s)}$-oscillations \eqref{eq:BsBd} and therefore
does not indicate signs of NP. It has to be mentioned that not only calculational uncertainties
but also NP could cancel in the ratios considered. Therefore inspection of individual branching ratios is necessary and in particular  an assessement of power corrections
becomes indispensable \cite{BZ06c}.

\section*{Addendum to v2}
Please note that in the arXiv version v1 
we used the BaBar bound  \cite{BaBar} $R_{\rm exp}^{\rm BaBar} < 0.029 \,\, {\rm
  at}\,\, 90\% \,\, {\rm CL}$,  
which was combined, by HFAG, with the Belle measurement to 
$R_{\rm exp}^{\rm HFAG}= 0.024\pm 0.006$ and resulted in $
\left| V_{td}/ V_{ts} \right|_{B\to V\gamma}^{\rm HFAG}  =  
0.179 \pm 0.014({\rm th}) \pm 0.022 ({\rm exp})$. These values have
changed with the BaBar measurement  ~\cite{Babar} presented at ICHEP 2006. The corresponding 
new results for $|V_{td}/V_{ts}|$
is given in \eqref{Rtrad}.

\subsection*{Acknowledgements}
R.Z.\ is grateful to the organizers of CAQCD06 for hospitality and the opportunity to
present this work.

\end{document}